\documentclass[12pt]{article}
\usepackage{amsmath,slett,epsfig,macros,cite}
\usepackage{macros_static}

\begin{document}

\begin{titlepage}

\begin{flushright}
DESY 03-087\\
HU-EP-03/39\\
MS-TP-03-7\\
SFB/CPP-03-15
\end{flushright}

\vskip 0.5cm
\begin{center}
{\Large\bf 
Lattice HQET with exponentially 
improved statistical precision\\[0.5ex] 
}
\end{center}
\vskip 0.7cm
\vbox{
\centerline{
\epsfxsize=2.5 true cm
\epsfbox{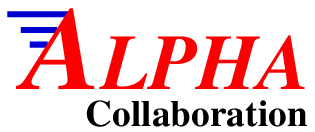}}
}
\vskip 0.5cm
\begin{center}
{% \large
Michele Della Morte$^{\scriptscriptstyle a}$,
Stephan D\"urr$^{\scriptscriptstyle a}$,
Jochen Heitger$^{\scriptscriptstyle b}$,
Heiko Molke$^{\scriptscriptstyle a}$,\\
Juri Rolf$^{\scriptscriptstyle c}$,
Andrea Shindler$^{\scriptscriptstyle d}$ and  
Rainer Sommer$^{\scriptscriptstyle a}$
}
\vskip 0.5cm
{\small
$^{\scriptstyle a}$
DESY, Zeuthen, Germany
\vskip 1.0ex
$^{\scriptstyle b}$
Institut f\"ur Theoretische Physik, Universit\"at M\"unster,
M\"unster, Germany
\vskip 1.0ex
$^{\scriptstyle c}$ 
Institut f\"ur Physik, Humboldt Universit\"at, 
Berlin, Germany
\vskip 1.0ex
$^{\scriptstyle d}$ NIC, Zeuthen, Germany
}
\vskip 0.875cm
{\bf Abstract}
\vskip 0.1ex
\end{center}
{\small
We introduce an alternative discretization for static quarks
on the lattice retaining the $\Oa$-improvement
properties of the Eichten-Hill action. In this formulation,
statistical fluctuations are reduced by a factor which grows
exponentially with Euclidean time, $x_0$.
For the first time, B-meson correlation functions
are computed with good statistical precision in the static approximation 
for $x_0>1\,\fm$.
At lattice spacings $a \approx 0.1\, \fm,\, 0.08\, \fm,\, 0.07\,\fm$,
the B$_{\rm s}$-meson decay constant is 
determined in the combined static and quenched 
approximation. A correction due to the finite 
mass of the b-quark is estimated by interpolating between
the static result and a recent determination of $\Fds$.
}


\vskip 0.29cm
\vfill

\begin{center}
\today %  February 2003
\end{center}

\eject
\vfill
\eject

\end{titlepage}

\newcommand{\ssect}[1]{\vspace{0.5cm}\noindent {\bf  #1.}}

%\section{The new actions}
\ssect{1} 
B-physics matrix elements such as the B-meson decay constant
$\fb$ are obtained from lattice correlation
functions at large Euclidean time. Considerable interest lies in the
treatment of the b-quark in the leading order of HQET, the static approximation
\cite{stat:eichten,stat:eichhill1}: in this
framework non-perturbative renormalization can be performed,
the continuum limit exists and also $1/\mbeauty$ corrections can in principle
be taken into 
account~\cite{zastat:pap1,lat01:rainer,lat02:rainer,lat02:jochen}.

Progress along this line has been hampered by large statistical errors
in the static approximation. In particular it has been
observed~\cite{stat:hashi} that the errors of a B-meson correlation function
roughly grow as
\be \label{e_SN}
  R_{\rm NS}\equiv {\rm noise \over \rm signal} % \equiv {N \over S} 
  \propto \exp\left(x_0\,\Delta\right)\,, \; \Delta=\Estat-\mpi\,,
\ee
where $\Estat$ is the ground state energy 
of a B-meson in the static approximation with the
Eichten-Hill action\footnote{For a more precise definition of
the theory and for any unexplained notation we refer to 
\cite{zastat:pap1}.}, 
\bes
  \label{e_LatLag}
  \Sstat^{\rm EH} &=& a^4  
            \sum_x \heavyb(x) D_0 \heavy(x)\,, \\
     D_0\heavy(x) &=& 
 {1\over a}\,[\heavy(x) - U^\dagger(x-a\hat{0},0)\heavy(x-a\hat{0})]\,, 
\label{e_EH}
\ees
for the static quark \cite{stat:eichhill1}. 
\Eq{e_SN} is problematic because the requirement $R_{\rm NS} \ll 1$
is satisfied only for $x_0$ of the order of $\Delta^{-1}$ and this
time interval shrinks rapidly to zero in the continuum limit $a\to0$ where
$\Estat \sim e_1 \times g_0^2 /a$ with some 
number $e_1$. In the attempt to eliminate the 
discretization errors by reducing the lattice spacing, $a$, 
one is then limited 
more and more by unwanted contaminations by higher energy states and
it has been very difficult to compute matrix elements in
the static approximation 
\cite{stat:eichten,fbstat:old1,fbstat:old2,stat:fnal2,reviews:beauty}.
Since the exponent in \eq{e_SN} is dominated by a divergent term,
it is plausible that one may reduce it by changing the discretization.
Here we will demonstrate that this is indeed possible while {\em remaining
with roughly the same discretization errors}. 

In \cite{zastat:pap1} it has been shown that energy differences
computed with the action \eq{e_LatLag} are $\Oa$-improved
if the relativistic sector (light quarks and gluons) is $\Oa$-improved.
Furthermore, apart from the usual mass dependent
factor, $1+\bastat a\mq$, the static axial current,
\bes
 \Astat(x)=\lightb(x)\gamma_0 \gamma_5\heavy(x)\,,
\ees
is on-shell $\Oa$-improved after adding only one correction term,
\bes \label{e_aimpr}
  \Astatimpr= \Astat+a\castat\delta\Astat\,, \quad 
  \delta\Astat(x) = \lightb(x)\gamma_j\gamma_5
  {\lnab{j}+\lnabstar{j} \over 2}\heavy(x)\,.
\ees
We want to retain these properties of the theory. They are guaranteed
if the lattice Lagrangian is invariant under the following
symmetry transformations (we do not list the usual 
ones such as parity and cubic invariance) \cite{zastat:pap1}.
\bi
 \item[i)] Heavy quark spin symmetry:
        \bes
          \label{e_spin}
          \heavy\longrightarrow {\cal V}\heavy\,,
          \qquad \heavyb\longrightarrow\heavyb {\cal V}^{-1}\,,\quad
          {\rm with} \quad
          {\cal V}=\exp(-i\phi_i \epsilon_{ijk}\sigma_{jk})\,. 
        \ees
 \item[ii)] Local conservation of heavy quark flavor number:
        \be
          \label{e_quarknumber}
           \heavy\longrightarrow \rme^{i\eta(\vecx)}\,\heavy\,,
          \qquad \heavyb\longrightarrow\heavyb \rme^{-i\eta(\vecx)}.
        \ee
\ei
Keeping these symmetries intact, there is little freedom to modify 
the action.
We may, however, alter the way the gauge fields
enter the discretized covariant derivative, $D_0$. 
To this end we choose
\be \label{e_newd0}
  D_0 \heavy(x) = 
  {1\over a}\,[\heavy(x) - W^\dagger(x-a\hat{0},0)\,\heavy(x-a\hat{0})]\,,
\ee
with $W(x,0)$ a generalized gauge parallel transporter
with the gauge transformation properties of $U(x,0)$.
In particular we take $W(x,0)$ to be a function of the 
link variables in the neighborhood of $x$, which is invariant  
under spatial cubic rotations and does have the correct
classical continuum limit such that $D_0=\partial_{0}+A_0+\rmO(a^2)$. 
This is enough to ensure that the
universality class as well as   
$\Oa$-improvement are unchanged in comparison to \eq{e_EH}. %the Eichten-Hill action. 
Since we expect the size of remaining higher order
lattice artifacts to be moderate if one keeps the action
rather local, we here consider only choices where $W(x,0)$ is a function
of gauge fields in the immediate neighborhood of $x, x+a\hat{0}$.
We choose
\bes
  W_{\rm S}(x,0) &=& V(x,0)\,\left[{g_0^{2} \over 5}+ 
                \Big({1\over 3}\tr V^\dagger(x,0) V(x,0)
                \Big)^{1/2}\right]^{-1}
  \,,\label{e_sx}\\
  W_{\rm A}(x,0) &=& V(x,0)
  \,,\label{e_ape}\\
  W_{\rm HYP}(x,0) &=& V_{\rm HYP}(x,0)\,,  \label{e_hyp}
\ees 
where 
\bes \label{e_staples}
  V(x,0) = {1\over6} \sum_{j=1}^3 &[&
           U(x,j)U(x+a\hat{j},0)U^{\dagger}(x+a\hat{0},j) \nonumber \\[-1ex]
           &&+\, 
           U^{\dagger}(x-a\hat{j},j)U(x-a\hat{j},0)
                         U(x+a\hat{0}-a\hat{j},j)\,\,\,\,] \,,
\ees
and where the so-called HYP-link, 
$V_{\rm HYP}(x,0)$, is a function of the gauge links 
located within a hypercube \cite{HYP,HYP:pot}.
In the latter case we take the parameters $\alpha_1=0.75$, $\alpha_2=0.6$, 
$\alpha_3=0.3$ \cite{HYP}. 
The choices (\ref{e_sx}) -- (\ref{e_hyp}) will be motivated 
further in \cite{stat:actions}. It is worth pointing out that a 
covariant derivative of the general type 
used above has first been introduced in \cite{fatlinks}.
In this reference it was considered for the Kogut-Susskind action
for relativistic quarks and with a different motivation.

%\section{Scaling}
\ssect{2} 
Next we have to study the scaling behaviour of observables computed 
with the actions $\Sstat^{\rm S},\, \Sstat^{\rm A},\, \Sstat^{\rm HYP}$
 which are obtained by
inserting $W_{\rm S},\,W_{\rm A},\,W_{\rm HYP}$ 
into eqs.~(\ref{e_newd0}) and (\ref{e_LatLag}). In~\cite{stat:actions}
this scaling behaviour is analyzed in depth for various observables
and various choices for the static action in perturbation theory and
non-perturbatively. Here we will present only one example.  The necessity
of such an investigation can be underlined by the following
consideration.

The static potential can be seen as an energy for a static
quark with action $\Sstat$ and an antiquark with the corresponding
$\Sastat$ \cite{pot:intermed}. Hence, the static force 
is one indicator for the scaling behavior of these actions. In
\cite{HYP:pot}, rather large $a^2$-effects have been seen in the
short-distance force for $\Sstat^{\rm HYP}$ and 
$\Sastat^{\rm HYP}$. 

One may therefore worry about large $a$-effects,
in particular  in correlation functions of 
the static-light axial current, where static and light quarks 
propagate also close to each other.  
With the new actions, $\Astat$ is $\Oa$-improved
once~\cite{stat:actions}
\bes
  \castat &=& -0.08237\,g_0^2 + \rmO(g_0^4) \,,\quad \mbox{for}\;\Sstat=\Sstat^{\rm EH}\,, \\
  \castat &=& 0.0072(4)\,g_0^2 + \rmO(g_0^4) \,,\quad \mbox{for}\;\Sstat=\Sstat^{\rm S},\Sstat^{\rm A}\,, \\
  \castat &=& 0.0385(37)\,g_0^2 + \rmO(g_0^4) \,,\quad \mbox{for}\;\Sstat=\Sstat^{\rm HYP} \, ,
\ees
is set in \eq{e_aimpr}. The improvement coefficient $b_{\rm A}^{\rm
  stat}$ is set to its 
tree--level value  $b_{\rm A}^{\rm stat}=1/2$ in this work.

\begin{figure}[htb]
\hspace{0cm}
\vspace{-1.0cm}

\centerline{
\psfig{file=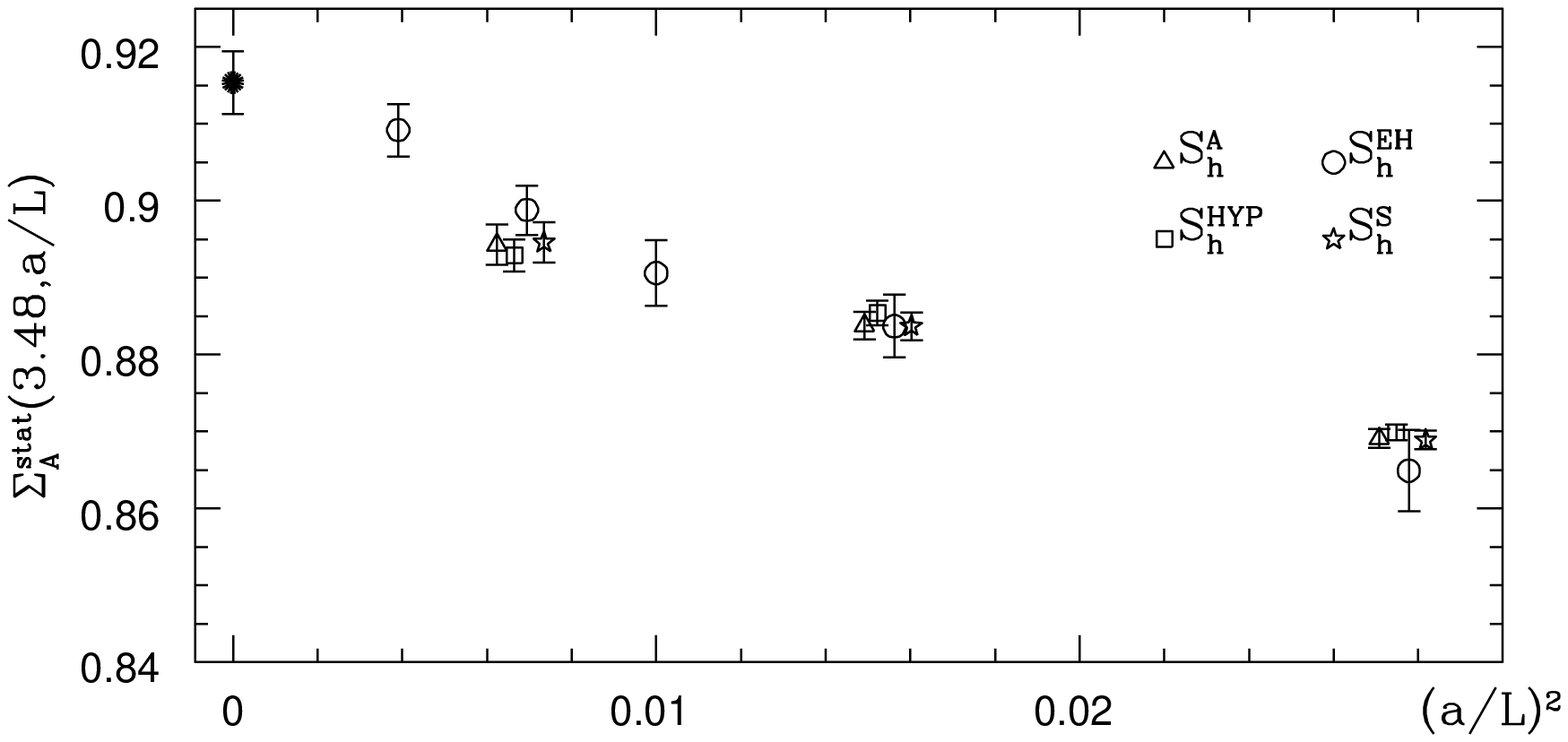,width=15cm,angle=0}
}
\vspace{-7.4cm}
\caption{\footnotesize 
         The step scaling function $\SigmaAstat(3.48,a/L)$
        for different choices of the action $\Sstat$.
        Results for $\Sstat^{\rm EH}$ 
        were extrapolated to $\SigmaAstat(3.48,0)$ 
        \protect\cite{zastat:pap3}
        ($\bullet$). In all cases $\castat$ from 1-loop perturbation
        theory is used, which is sufficient since
         $\SigmaAstat(3.48,a/L)$ does not depend
        very sensitively on this improvement coefficient. 
        For  $\Sstat^{\rm A}$, $\Sstat^{\rm S}$ and $\Sstat^{\rm HYP}$
        points have been displaced on the horizontal axis for clarity.
\label{f_ssf}}
\end{figure}

We consider now a step scaling function, $\SigmaAstat$,
which gives the change of the renormalized static axial current 
in a \SF (SF) scheme \cite{zastat:pap3}, 
when the renormalization scale is changed from
$\mu=1/L$ to $\mu=1/(2L)$. Its continuum 
limit is known for a few values
of $L$ \cite{zastat:pap3}. This quantity is thus a good observable to 
search for 
$a$-effects. In \fig{f_ssf} we show $\SigmaAstat(3.48,a/L)$, where the first
%%%%%%%%%%%%%%%%%%%%%%%%%%%%%FIGURE%%%%%%%%%%%%%%%%%%%%%%%%%%%%%%%%%%%
%
%%%%%%%%%%%%%%%%%%%%%%%%%%%%%FIGURE%%%%%%%%%%%%%%%%%%%%%%%%%%%%%%%%%%%
argument parameterizes $L$ in terms of the SF-coupling $\gbar^2(L)=3.48$. 
$\Oa$-improvement is employed as in \cite{zastat:pap3} but
we consider the different actions for the static quark introduced above. 
All of them lead to $\SigmaAstat(3.48,a/L)$ at finite $a/L$ differing 
from the continuum limit 
by about the same amount. Supported also by further such studies
\cite{stat:actions}, we conclude that within the set of actions
studied none is particularly distinguished by its scaling behavior. 

%%%%%%%%%%%%%%%%%%%%%%%%%%%%%%%%%%%%%%%%%%%%%%%%%%%%%%%%%%%%%%%%%%%%%%%%%

\begin{figure}[htb]
\hspace{0cm}
\vspace{-0.0cm}

\centerline{
\psfig{file=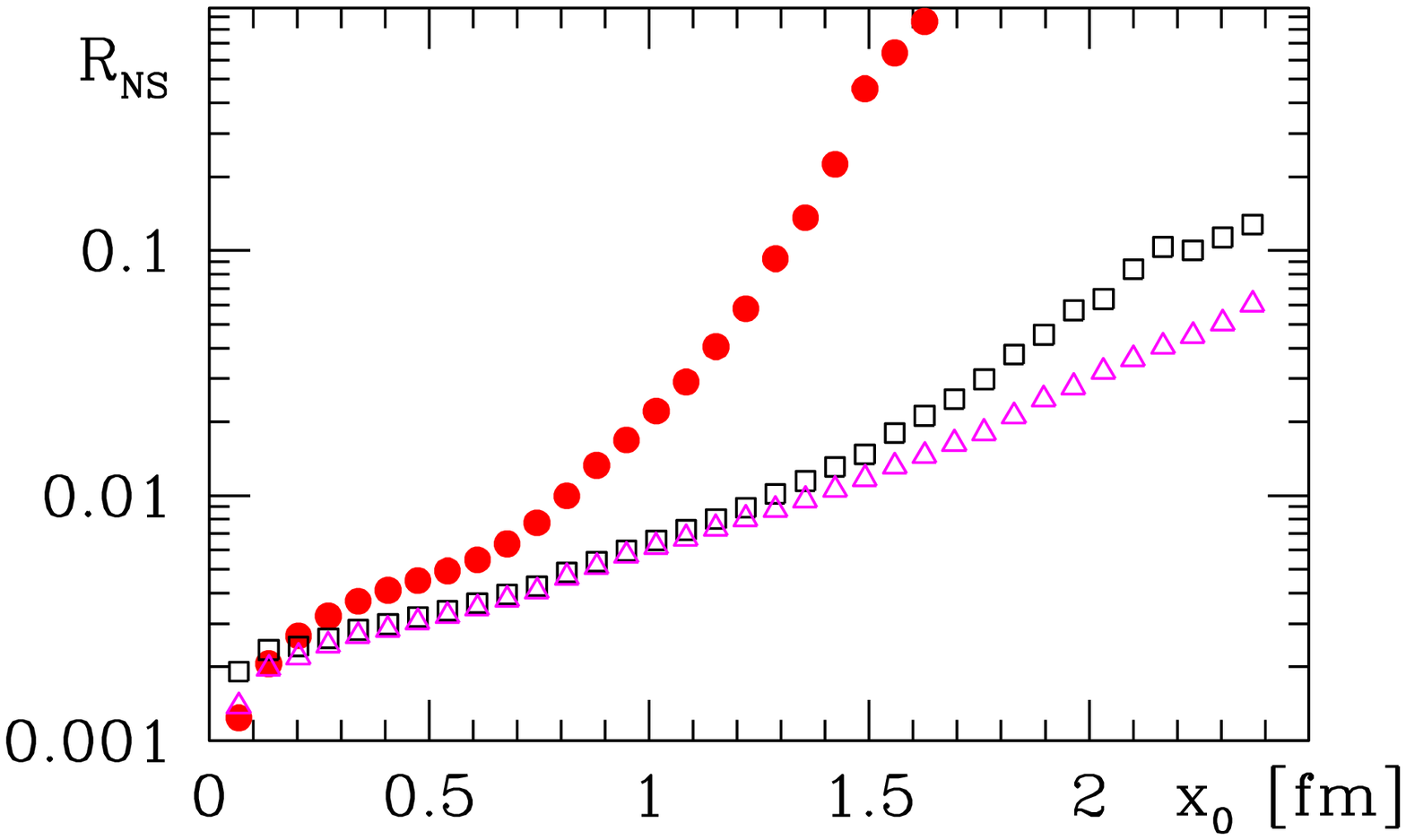,width=8cm,angle=0}
}
\vspace{-0.0cm}
\caption{\footnotesize 
         The ratio $R_{\rm NS}$, \eq{e_SN}, for the correlation function
         $\fastat$ for a statistics of 2500 measurements. 
         Circles refer to  $\Sstat^{\rm EH}$ while squares and triangles 
         to $\Sstat^{\rm S}$ and $\Sstat^{\rm HYP}$, respectively.
         $\Sstat^{\rm A}$ behaves like  $\Sstat^{\rm S}$.
         Physical units are set by using
         $\rnod=0.5\,\fm$ \protect\cite{pot:r0,pot:r0_SU3}.
\label{f_rns}}
\end{figure}

\ssect{3} Let us now demonstrate that the statistical errors
at large Euclidean time are reduced by the choices 
eqs.~(\ref{e_sx}) -- (\ref{e_hyp}). 
As a B-meson correlation function we choose
\be \label{e_fastat}
 \fastat(x_0,\omega) 
        = - \frac12 \langle \Astatimpr(x) \, \op{}(\omega) \rangle\,,
        \quad \op{}(\omega) = 
        {a^6\over L^3} \sum_{\vecy,\vecz} \zetahb(\vecy) \gamma_5  
        \omega(\vecy-\vecz) \zetal(\vecz) \,,
\ee
defined in the \SF with $T=3L/2$, $L/a=24$, $\beta=6/g_0^2=6.2$ and
a vanishing background field \cite{zastat:pap1}. 
Here, as a novelty compared to previous applications, a wave function 
$\omega(\vecx)$ is introduced to construct an interpolating B-meson field 
in terms of the boundary quark fields $\zetal$ and
$\zetahb$. It may be exploited to reduce the contribution
of excited B-meson states to the correlation function, but this does not
concern us yet. At the moment we simply 
consider $\omega(\vecx)=1$ and form the ratio
$R_{\rm NS}$, \eq{e_SN}, for the different actions. From now on
we set the light quark mass to the {\em strange quark mass},
taken from \cite{mbar:pap3} following exactly \cite{mbar:charm1} 
concerning the technical details.\footnote{Of course, these details matter 
only before taking the continuum limit.} 
\Fig{f_rns} shows that in all 
%%%%%%%%%%%%%%%%%%%%%%%%%%%%%FIGURE%%%%%%%%%%%%%%%%%%%%%%%%%%%%%%%%%%%
%
%%%%%%%%%%%%%%%%%%%%%%%%%%%%%FIGURE%%%%%%%%%%%%%%%%%%%%%%%%%%%%%%%%%%%
cases $R_{\rm NS}$ grows exponentially with $x_0$. For the Eichten-Hill 
action, also the effective 
coefficient $\Delta$, describing the growth
for $x_0=1\,\fm - 2\,\fm$, is {\em roughly} 
given by $\Estat-\mpi$ in agreement with \eq{e_SN}, 
while for the other actions this is not
the case. Most importantly for the other actions,  $\Delta$ is 
reduced by a factor around 4,
and with the statistics in our example a distance 
of $x_0 \approx 2\,\fm$ is reached with $\Sstat^{\rm HYP}$ 
if one requires $R_{\rm NS}\leq 2\%$. 
The actions 
$\Sstat^{\rm A},\Sstat^{\rm S}$ behave only slightly worse.

\begin{figure}[htb]
\hspace{0cm}
\vspace{-3.5cm}

\centerline{
\psfig{file=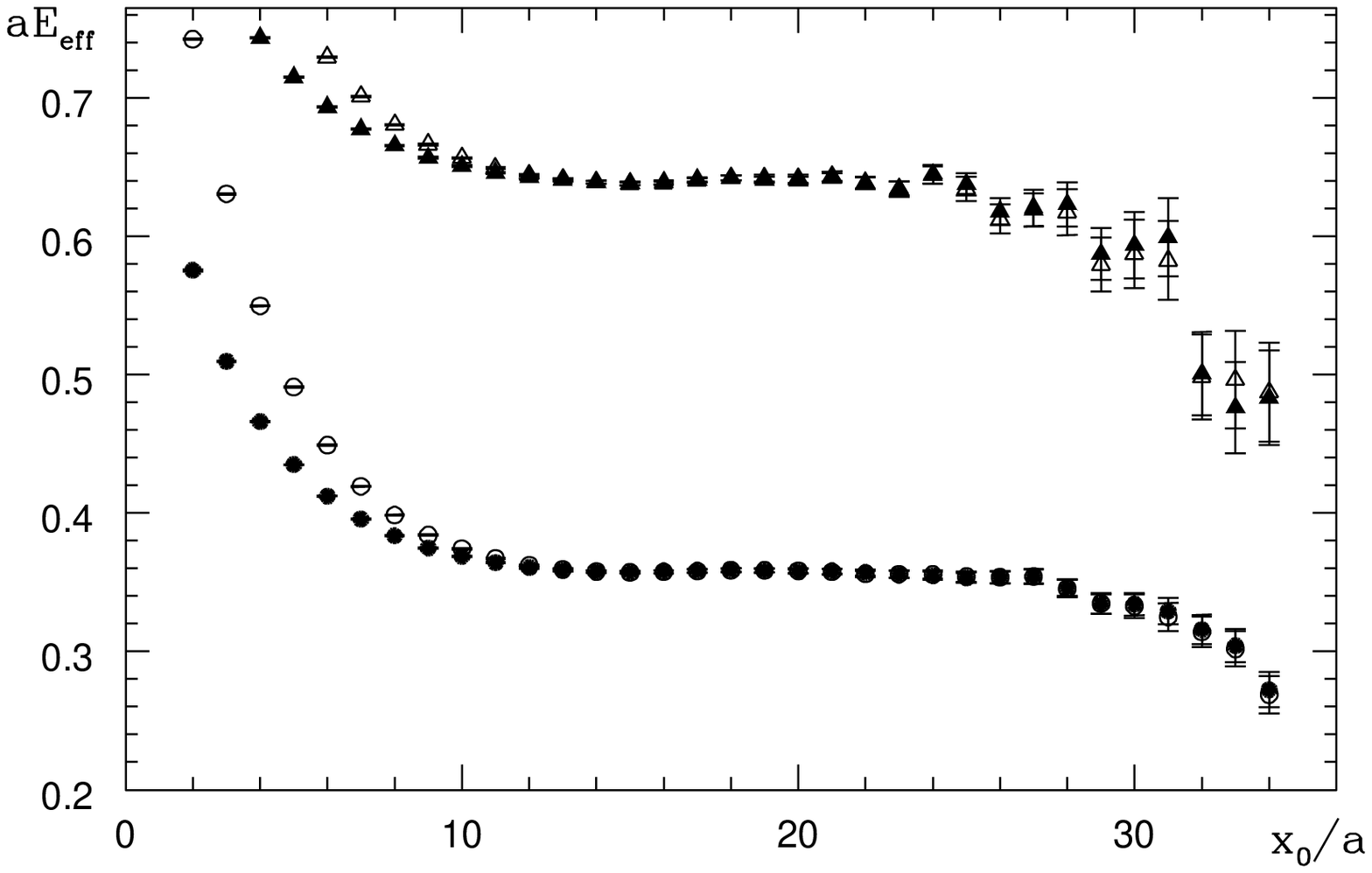,width=14.cm,angle=0}
}
\vspace{-2.7cm}
\caption{\footnotesize 
         Effective energies for wave functions $\Omega_1$ 
         (open symbols) and $\Omega_2$ (filled symbols) using 
         $\Sstat^{\rm HYP}$ (circles) and $\Sstat^{\rm S}$ (triangles).
         Results refer to a $24^3 \times 36$ lattice, $\beta=6.2$.
\label{f_plateau}}
\end{figure}

\ssect{4} This reduction of statistical errors enables
us to choose $\omega(\vecx)$ such that a long and precise
plateau is visible in the effective energy,
\be
  E_{\rm eff}(x_0,\omega) = 
        \ln\left[\fastat(x_0-a,\omega)/\fastat(x_0+a,\omega)\right]/(2a)\,,
\ee 
as shown in \fig{f_plateau}.
%%%%%%%%%%%%%%%%%%%%%%%%%%%%%FIGURE%%%%%%%%%%%%%%%%%%%%%%%%%%%%%%%%%%%
%
%%%%%%%%%%%%%%%%%%%%%%%%%%%%%FIGURE%%%%%%%%%%%%%%%%%%%%%%%%%%%%%%%%%%%
Neither position nor length of the plateau depend sensitively on the
details of $\omega$, as long as it is chosen such that the first
excited state in the B-meson channel is canceled to a good
approximation. For the figure as well as for the following, we have
chosen $\omega \in \{\Omega_1,\Omega_2\}$ with 
\bes \Omega_1 &=&
\omega_1 + \alpha \omega_3\,, \quad
\Omega_2   =  \omega_2 + \alpha' \omega_4\,, \\
\omega_i(\vecx) &=& N_i^{-1} \sum_{{\bf n}\in{\bf Z}^3}
\overline{\omega}_i(|\vecx-{\bf n}L|)\,,\;i=1,2,3\,,
\nonumber \\
\overline{\omega}_1(r) &=& r_0^{-3/2}\,\rme^{-r/a_0}\,,\quad
\overline{\omega}_2(r) = r_0^{-3/2}\,\rme^{-r/(2a_0)}\,,\quad
\overline{\omega}_3(r) =  r_0^{-5/2}\,r\,\rme^{-r/(2a_0)} \,,\nonumber \\
\omega_4(\vecx) &=& L^{-3/2}\,, 
\ees 
where $a_0=0.1863\,r_0$ and the
(dimensionless) coefficients $N_i$ are chosen such that $a^3\sum_{\bf
  x} \omega_i^2({\bf x})=1$.  The B-meson decay constant is then
obtained from the renormalization group invariant matrix
element~\cite{mbar:pap2} 
\be \label{e_feff} 
\Phirgi(x_0) = -
Z_{\rm RGI}\,(1+ b_{\rm A}^{\rm stat} am_{\rm q}) \,2 L^{3/2} {
{\fastat(x_0)}\over{\sqrt{f_1(T',\omega)}} }\,\rme^{(x_0-T'/2)
E_{\rm eff}(x_0)} 
\ee 
of the static axial current, where 
\bes
f_1(T,\omega) = 
-\frac12 \langle \op{}'(\omega) \op{}(\omega) \rangle\,,\quad 
\op{}'(\omega) = 
{a^6 \over L^3} \sum_{\vecy,\vecz} 
\zetalbprime(\vecy) \gamma_5\, \omega(\vecy-\vecz) \zetahprime(\vecz)\,.  
\ees
The renormalization factor, $Z_{\rm RGI}$, relates the bare matrix element 
to the renormalization group invariant one \cite{zastat:pap3}.
Its regularization dependent part is computed exactly as in that reference, 
but for the new actions.
In \tab{t_feff} we give results for $\Phirgi(x_0)$
for three values of the lattice spacing and selected choices of
$T,T',x_0$, highlighting what we selected for further analysis.
These numbers do not change significantly if we vary 
the improvement coefficients $\castat$ and $b_{\rm A}^{\rm stat}$,
which are known only in perturbation theory,
by factors of two.
We thus extrapolate our results quadratically in the lattice spacing and 
arrive at our estimate for the continuum limit
\be \label{e_fbstat}
  r_0^{3/2}\Phirgi = 1.78(13) \,.
\ee
%%%%%%%%%%%%%%%%%%%%%%%%%%%%%%%%%%%%%%%%%%%%%%%%%%%%%%%%%%%%%%%%%%%%
\begin{table}[htb]
\centering
\begin{tabular}{cccccccccc}
\hline
  & & & & & & $\Omega_1$ & $\Omega_2$ \\
  $\beta$   &   $a\,[{\rm fm}]$  &   $L/a$   &   $T/a$   &   $T'/a$  &  $x_0/a $  &
  \multicolumn{2}{c}{$r_0^{3/2}\Phirgi$} & $\alpha$ & $\alpha'$ \\
\hline
  6.0  & 0.093 & 16 & 24 & 24 & 12 & 1.830(31) & {\bf 1.832(30)}
& 0.278 & $-0.200$ \\ [1ex]
%       &      &    &    &    &     &   &  \\[1ex]
  6.0  & 0.093 & 16 & 24 & 20 & 12 & 1.847(18) & 1.830(17)
& 0.278 & $-0.200$ \\[1ex]
%       &      &    &    &    &     & ... &  \\[1ex]
  6.0  & 0.093 & 16 & 24 & 24 & 10 & 1.818(31) & 1.828(30)
& 0.278 & $-0.200$ \\[1ex]
  6.0  & 0.093 & 16 & 24 & 20 & 10 & 1.851(17) & 1.829(17)
& 0.278 & $-0.200$ \\[1ex]
\hline
  6.1  & 0.079 & 24 & 30 & 30 & 15 & 1.864(56) & {\bf 1.858(52)}
& 0.756 & 0.022 \\[1ex]
  6.1  & 0.079 & 24 & 30 & 30 & 12 & 1.850(56) & 1.846(52)
& 0.756 & 0.022 \\[1ex]
\hline
  6.2  & 0.068 & 24 & 36 & 36 & 18 & 1.724(78) & {\bf 1.760(75)}
& 0.351  & $-0.176$ \\[1ex]
  6.2  & 0.068 & 24 & 36 & 36 & 15 & 1.726(78) & 1.763(76)
& 0.351  & $-0.176$ \\[1ex]
\hline \\[-2.0ex]
\end{tabular}
\caption{\footnotesize{
 Decay constant in static approximation.
}}\label{t_feff}
\end{table}

\ssect{5} The result \eq{e_fbstat} 
may be used to compute $\Fbs$ by taking account of the mass dependent
function~\cite{zastat:pap3} $C_{\text{PS}}(M_{\text{b}}/\Lambda_{\MSbar}) =
\Fb\sqrt{m_{\text{B}}}/\Phirgi = 1.22(3) $,
evaluated using the 3-loop anomalous dimension~\cite{ChetGrozin} and
the associated matching coefficient between HQET and QCD
\cite{BroadhGrozinII}.  $M_{\rm b}$ denotes the renormalization group
invariant b-quark mass~\cite{lat01:rainer,lat02:rainer}.  With this we
arrive at $r_0\Fbs^{\text{stat}} = 0.57(4)$.  A correction due to the
finite mass of the b-quark can be computed by connecting the static
result~\eq{e_fbstat} and \be r_0^{3/2} {{F_{\rm D_s}\sqrt{m_{\rm
        D_s}}} \over{C_{\rm PS}(M_{\rm c}/\Lambda_{\MSbar})}} =
1.33(7) \ee by a linear interpolation in the inverse meson mass. Here
we have used recent computations of the D$_{\text{s}}$-meson decay
constant~\cite{fds:JR03} and of the charm quark
mass~\cite{mbar:charm1}.  In this way we obtain \be
\label{e_fbstat_qcd} r_0 \Fbs = 0.52(3) \to \Fbs = 205(12)\,\MeV
\;\mbox{with}\; r_0=0.5\,\fm\,.  \ee Conservatively, we may attribute
an additional $\approx 5\,\%$ uncertainty to the fit ansatz used, but
our personal estimate is that this error is significantly smaller and
it will soon be quantified~\cite{alphaprep}.

One should remember that \eq{e_fbstat_qcd} refers to the quenched 
approximation and as in \cite{fds:JR03} a $12\,\%$ scale ambiguity
may be estimated from the slope of the linear interpolation.

\ssect{6}
An interesting point is that 
the potential in full QCD may be computed replacing the 
time-like links in the Wilson loop (or Polyakov loops)
by the different $W_i$ introduced above. In particular the
``HYP-link potential'' \cite{HYP:pot} may be used. Depending
on which $W_i$ is chosen, 
the static potentials differ from each other, 
but all of them approach the continuum limit with
$\rmO(a^2)$ corrections if the action used for the dynamical
fermions is $\Oa$-improved. This
property follows from the considerations of
\cite{pot:intermed} applied to the static actions introduced above, 
which satisfy all the necessary requirements. This virtue of
e.g. the HYP-link potential was not obvious before. 
Using it, better precision can be reached 
and some signs of string breaking \cite{lat99:klaus} may become visible.

\ssect{7} To summarize, we have shown that a modification of the 
Eichten-Hill static action can be found which keeps lattice artifacts 
in heavy-light correlation functions moderate but reduces statistical errors
to a level making the region $x_0>1.5\,\fm$ accessible. Furthermore,
the new action can be used without change for dynamical fermions
and also to compute the static potential with dynamical fermions. 
As a demonstration of the usefulness of this reduction of statistical
errors, we have computed $\Fbs$ in the quenched approximation,
by joining the continuum limit of the static approximation 
estimated with the new action with the previously determined
continuum limit of $\Fds$ by means of a linear interpolation. This procedure
can systematically be improved by computing 1) the mass dependence
around $\mcharm$, 2) the $1/m$ corrections to the static limit
and 3) repeating the whole analysis with dynamical fermions. 
Work along these lines is in progress and a more detailed investigation
of the properties of various static quark actions is in preparation. 

\ssect{Acknowledgments} 
We are grateful to M. L\"uscher and F. Knechtli for comments on
the manuscript.
We thank DESY for allocating computer time on the APEmille computers
at DESY Zeuthen to this project and the APE-group for its valuable
help.
This work is also supported by the EU IHP Network on
Hadron Phenomenology from Lattice QCD under grant HPRN-CT-2000-00145
and by the Deutsche Forschungsgemeinschaft in the SFB/TR~09.

  \bibliographystyle{h-elsevier3}

\begin{thebibliography}{99}

\bibitem{stat:eichten}
E. Eichten,
\newblock Talk delivered at the Int. Sympos. of Field Theory on the Lattice,
  Seillac, France, Sep 28 - Oct 2, 1987.

\bibitem{stat:eichhill1}
E. Eichten and B. Hill,
\newblock Phys. Lett. B234 (1990) 511.
%%CITATION = PHLTA,B234,511;%%

\bibitem{zastat:pap1}
ALPHA, M. Kurth and R. Sommer,
\newblock Nucl. Phys. B597 (2001) 488, hep-lat/0007002.
%%CITATION = NUPHA,B597,488;%%

\bibitem{lat01:rainer}
ALPHA, J. Heitger and R. Sommer,
\newblock Nucl. Phys. Proc. Suppl. 106 (2002) 358, hep-lat/0110016.
%%CITATION = HEP-LAT 0110016;%%

\bibitem{lat02:rainer}
R. Sommer,
\newblock (2002), hep-lat/0209162.
%%CITATION = HEP-LAT 0209162;%%

\bibitem{lat02:jochen}
J. Heitger, M. Kurth and R. Sommer,
\newblock (2002), hep-lat/0209078.
%%CITATION = HEP-LAT 0209078;%%

\bibitem{stat:hashi}
S. Hashimoto,
\newblock Phys. Rev. D50 (1994) 4639, hep-lat/9403028.
%%CITATION = HEP-LAT 9403028;%%

\bibitem{fbstat:old1}
C.R. Allton et~al.,
\newblock Nucl. Phys. B349 (1991) 598.
%%CITATION = NUPHA,B349,598;%%

\bibitem{fbstat:old2}
C. Alexandrou et~al.,
\newblock Phys. Lett. B256 (1991) 60.
%%CITATION = PHLTA,B256,60;%%

\bibitem{stat:fnal2}
A. Duncan et~al.,
\newblock Phys. Rev. D51 (1995) 5101, hep-lat/9407025.
%%CITATION = PHRVA,D51,5101;%%

\bibitem{reviews:beauty}
R. Sommer,
\newblock Phys. Rept. 275 (1996) 1, hep-lat/9401037.

\bibitem{HYP}
A. Hasenfratz and F. Knechtli,
\newblock Phys. Rev. D64 (2001) 034504, hep-lat/0103029.
%%CITATION = HEP-LAT 0103029;%%

\bibitem{HYP:pot}
A. Hasenfratz, R. Hoffmann and F. Knechtli,
\newblock Nucl. Phys. Proc. Suppl. 106 (2002) 418, hep-lat/0110168.
%%CITATION = HEP-LAT 0110168;%%

\bibitem{stat:actions}
M. Della~Morte, A. Shindler and R. Sommer,
\newblock in preparation.

\bibitem{fatlinks}
T.~Blum et~al.,
%``Improving flavor symmetry in the Kogut-Susskind hadron spectrum,''
Phys. Rev. D55 (1997) 1133,
hep-lat/9609036.
%%CITATION = HEP-LAT 9609036;%%

\bibitem{pot:intermed}
S. Necco and R. Sommer,
\newblock Nucl. Phys. B622 (2002) 328, hep-lat/0108008.
%%CITATION = HEP-LAT 0108008;%%

\bibitem{zastat:pap3}
J. Heitger, M. Kurth and R. Sommer,
\newblock (2003), hep-lat/0302019.
%%CITATION = HEP-LAT 0302019;%%

\bibitem{mbar:pap3}
ALPHA, J. Garden et~al.,
\newblock Nucl. Phys. B571 (2000) 237, hep-lat/9906013.
%%CITATION = NUPHA,B571,237;%%

\bibitem{mbar:charm1}
ALPHA, J. Rolf and S. Sint,
\newblock JHEP 12 (2002) 007, hep-ph/0209255.
%%CITATION = HEP-PH 0209255;%%

\bibitem{pot:r0}
R. Sommer,
\newblock Nucl. Phys. B411 (1994) 839, hep-lat/9310022.

\bibitem{pot:r0_SU3}
ALPHA, M. Guagnelli, R. Sommer and H. Wittig,
\newblock Nucl. Phys. B535 (1998) 389, hep-lat/9806005.
%%CITATION = NUPHA,B535,389;%%

\bibitem{mbar:pap2}
ALPHA, M. Guagnelli et~al.,
\newblock Nucl. Phys. B560 (1999) 465, hep-lat/9903040.
%%CITATION = HEP-LAT 9903040;%%

\bibitem{ChetGrozin}
K.G. Chetyrkin and A.G. Grozin,
\newblock (2003), hep-ph/0303113.
%%CITATION = HEP-PH 0303113;%%

\bibitem{BroadhGrozinII}
D.J. Broadhurst and A.G. Grozin,
\newblock Phys. Rev. D52 (1995) 4082, hep-ph/9410240.
%%CITATION = HEP-PH 9410240;%%

\bibitem{fds:JR03}
ALPHA, A. J{\"u}ttner and J. Rolf,
\newblock Phys. Lett. B560 (2003) 59, hep-lat/0302016.
%%CITATION = HEP-LAT 0302016;%%

\bibitem{alphaprep}
ALPHA,
\newblock in preparation .

\bibitem{lat99:klaus}
K. Schilling,
\newblock Nucl. Phys. Proc. Suppl. 83 (2000) 140, hep-lat/9909152.
%%CITATION = HEP-LAT 9909152;%%

\end{thebibliography}
\end{document}